\documentclass[twocolumn,prl,preprintnumbers,%showpacs,
nofootinbib,superscriptaddress,amsmath]{revtex4}

\allowdisplaybreaks

\renewcommand{\d}{\mathrm{d}}
\newcommand{\I}{\mathrm{i}}
\newcommand{\e}{\mathrm{e}}
\newcommand{\p}{\partial}
\newcommand{\cK}{\mathcal{K}}
\newcommand{\zb}{{\bar{z}}}
\newcommand{\tab}{\quad\,}

\DeclareMathOperator{\re}{\mathrm{Re}}

\DeclareSymbolFont{AMSb}{U}{msb}{m}{n}
\DeclareMathSymbol{\fieldR}{\mathalpha}{AMSb}{"52}
\DeclareMathSymbol{\fieldZ}{\mathalpha}{AMSb}{"5A}

\begin{document} %%%%%%%%%%%%%%%%%%%%%%%%%%%%%%%%%%%%%%%%%%%%%%%%%%%%%%%

\preprint{ITP--UU--05/25}
\preprint{SPIN--05/19}
\preprint{FSU--TPI--05/05}
\preprint{hep-th/0506181}

\title{On de~Sitter Vacua in Type IIA Orientifold Compactifications}

\author{Frank Saueressig}

\email{F.S.Saueressig@phys.uu.nl}

\affiliation{Institute for Theoretical Physics \emph{and} Spinoza
Institute \\ Utrecht University, 3508 TD Utrecht, The Netherlands}

\author{Ulrich Theis}

\email{Ulrich.Theis@uni-jena.de}

\affiliation{Institute for Theoretical Physics,
Friedrich-Schiller-University Jena, \\ Max-Wien-Platz 1, D-07743 Jena,
Germany}

\author{Stefan Vandoren}

\email{S.Vandoren@phys.uu.nl}

\affiliation{Institute for Theoretical Physics \emph{and} Spinoza
Institute \\ Utrecht University, 3508 TD Utrecht, The Netherlands}

\begin{abstract}

This letter discusses the orientifold projection of the quantum
corrections to type IIA strings compactified on rigid Calabi-Yau
threefolds. It is shown that $N=2$ membrane instanton effects give a
holomorphic contribution to the superpotential, while the perturbative
corrections enter into the K\"ahler potential. At the level of the
scalar potential the corrections to the K\"ahler potential give rise to
a positive energy contribution similar to adding anti-$D3$-branes in the
KKLT scenario. This provides a natural mechanism to lift an AdS vacuum
to a meta-stable dS vacuum.

\end{abstract}

%\pacs{}
%\keywords{supersymmetry, orientifolds, de~Sitter vacua}

\maketitle

\section{Introduction} %%%%%%%%%%%%%%%%%%%%%%%%%%%%%%%%%%%%%%%%%%%%%%%%%

Solving the problem of moduli stabilization is of primary importance for
deriving (semi-)realistic vacua from string theory compactifications.
Building on the work of KKLT \cite{KKLT}, subsequent investigations,
mostly in the context of orientifold compactifications of the type IIB
string and its $F$-theory dual \cite{EGQ}, but also for the type IIA
string \cite{A,KKP,DGKT,CFI,DKPZ}, have given evidence that suitable
combinations of background fluxes and non-perturbative effects like
instantons are capable of fixing all string moduli in an AdS vacuum.
Astronomical observations, however, indicate that our observable
universe has a small positive cosmological constant \cite{CC},
suggesting that one should search for string theory vacua with such a
property, i.e., dS vacua. In the context of the type IIB string, KKLT
outlined the construction of such vacua by first using background
fluxes and non-perturbative effects to stabilize all moduli in an AdS
vacuum. Subsequently, a positive energy contribution in form of a
$\overline{D3}$-brane was added to lift the AdS vacuum to a meta-stable
dS vacuum. After stabilizing all moduli except for the volume modulus
$\sigma$ of the compactification their potential reads \cite{KKLT}:
 \begin{equation}\label{1.0}
  V_\mathrm{KKLT} = \frac{a A\, \e^{-a \sigma}}{2\, \sigma^2} \Big( W_0
  + \frac{a}{3}\, \sigma A\, \e^{-a \sigma} + A\, \e^{-a \sigma} \Big)
  + \frac{D}{\sigma^3}\ .
 \end{equation}
Here the terms in brackets originate from fluxes, instanton corrections
and gluino condensation, while the last term proportional to $D$
corresponds to the positive energy contribution which was added by hand.
The first term of the potential \eqref{1.0} vanishes when switching off
the non-perturbative effects (setting $A=0$). In this case the type IIB
potential (without $\overline{D3}$-brane contribution) is independent of
the volume modulus, which is the known no-scale structure of IIB
orientifold compactifications. Stabilizing all (geometric) moduli in
this framework then requires the inclusion of non-perturbative effects
as discussed above. Furthermore, for $D>0$ the parameters $a$, $A$,
$W_0$ can be chosen such that $V_\mathrm{KKLT}$ has a meta-stable dS
vacuum.

In this letter we focus on orientifold projections of type IIA string
theory compactified on a rigid Calabi-Yau threefold (CY$_3$). Here it is
possible to stabilize all (geometric) moduli in an AdS vacuum using
classical fluxes only \cite{DGKT,CFI}. To our knowledge, the uplift of
these vacua to meta-stable dS vacua has not yet been performed; the aim
of this letter is to derive a mechanism for such an uplift.

Starting from the perturbative and membrane instanton corrections to the
universal hypermultiplet arising from compactifications of the type IIA
string on a rigid CY$_3$ \cite{DSTV}, we perform the orientifold
projection along the lines of \cite{GL}. The result agrees with the
standard folklore that the K\"ahler potential (describing the projected
hypermultiplet sector) receives perturbative corrections, while the
leading instanton corrections yield a holomorphic contribution to the
$N=1$ superpotential. Secondly, we study the scalar potential arising
from a space-time filling 4-form flux together with the leading
instanton contributions. It will turn out that, at the level of the
scalar potential, the perturbative corrections to the K\"ahler potential
give rise to a positive energy contribution, analogous to adding an
$\overline{D3}$-brane in the potential \eqref{1.0}. Therefore, these
corrections provide a natural mechanism which can lift an AdS vacuum to
a meta-stable dS vacuum.\footnote{In the context of type IIB orientifold
compactifications a similar lifting due to $\alpha'$ corrections to the
K\"ahler potential has been discussed in \cite{BB}.}

\section{The N=1 orientifold model} %%%%%%%%%%%%%%%%%%%%%%%%%%%%%%%%%%%%

In the following we will consider $N=1$ supergravity coupled to a single
chiral multiplet. Our model describes a subsector of an orientifold
projection of type IIA strings on a rigid CY$_3$. Before the projection,
the $N=2$ supergravity effective action comprises the universal
hypermultiplet (containing the dilaton) and $h_{1,1}$ vector multiplets.
The combined scalar target space is the product of a (four-dimensional)
quaternion-K\"ahler (QK) space and a special K\"ahler manifold
parametrized by the complexified K\"ahler moduli.

After projecting to $N=1$, the QK space becomes a (two-dimensional)
K\"ahler space parametrized by a single chiral multiplet containing the
dilaton $\phi$ and RR scalar $v$. Classically, they define the complex
structure
 \begin{equation} \label{c-str}
  z = 2 \e^{\phi/2} + \I v\ .
 \end{equation}
In our conventions the asymptotic value of the dilaton is related to
the string coupling constant by $\e^{-\phi_\infty/2}=g_s$. The special
K\"ahler manifold remains special K\"ahler, but of lower dimension
\cite{GL}. Classically, the total scalar manifold retains the product
structure inherited from the $N=2$ theory. Consistency of the
orientifold projection requires that the tadpole for the RR 7-form
potential, arising from the presence of $O6$-planes, is suitably
canceled. This can be done by adding $D6$-branes \cite{DGKT}.
In the following, we consider the sector of the universal hypermultiplet
only, assuming that the K\"ahler moduli have already been stabilized
(presumably by 4-form fluxes $F_4$ \cite{DGKT}).

The potential for the chiral scalar $z$ is then given by
 \begin{equation} \label{1.1}
  V = \e^\cK \big[\, \cK^{z\zb} D_z W\, \bar{D}_\zb \bar{W} - 3
  |W|^2\, \big]\ ,
 \end{equation}
where $\cK(z,\zb)$ is the K\"ahler potential, $\cK^{z\zb}=(\p_z\p_\zb
\cK)^{-1}$ the inverse K\"ahler metric, $W(z)$ the holomorphic
superpotential, and $D_z W=\p_zW+W\p_z\cK$. We consider a K\"ahler
potential of the form
 \begin{equation} \label{1.2}
  \cK = -2 \ln \big[ (z + \zb)^2 - 16\, c \big]\ .
 \end{equation}
Here, the constant $c\in\fieldR$ describes a loop
correction, since it is suppressed by powers of $g_s$. Furthermore, we
consider the superpotential
 \begin{equation} \label{1.3}
  W = 8 \, W_0 + \bar{A}\, \e^{-z}\ ,
 \end{equation}
with two constant parameters $W_0$ and $A$. We can assume without loss
of generality that $W_0$ is real, while $A$ is taken to be complex. The
constant term in $W$ arises from background fluxes, while the
exponential term can be understood as a membrane instanton correction.

We now investigate the properties of the scalar potential $V(z,\zb)$
in the presence of a non-vanishing $c$. It turns out to be convenient
to express the potential in terms of 
 \begin{equation} \label{zpert}
  z = 2 \sqrt{r+c} + \I v\ ,
 \end{equation}
where we denote $r=\e^\phi$ and require $r+c\geq 0$. As we show below,
this is the one-loop corrected complex structure inherited from the
$N=2$ theory.\footnote{Quantum corrections to complex structures also 
appear in three-dimensional gauge theories with four supercharges 
\cite{BHOY,DTV}.} We can then compute the potential and find
 \begin{align} \label{1.4}
  V & = \frac{W_0^2}{4 \, r^2} + \frac{W_0}{8\, r^{3/2}}\, \big(
	\bar{A}\, \e^{-z} + A\, \e^{-\zb} \big) \notag \\[2pt]
  & \tab - \frac{ c\, W_0^2}{r^2 (r+2c)} + \dots\ ,
 \end{align}
where $r$ should be understood as a function of $z$ and $\zb$ via the
relation \eqref{zpert}. We have included only the leading instanton
contribution to $V$; subleading terms proportional to $\e^{-(z+\zb)}$,
which belong to the 2-instanton sector not taken into account in the
superpotential \eqref{1.3}, have been dropped for consistency. Moreover,
we have expanded the $r$-dependent prefactor of the 1-instanton term for
large $r>-c$ and kept only the leading power.

The last term in $V$ is induced by the constant $c$ appearing in the
K\"ahler potential \eqref{1.2}. Comparing to \eqref{1.0}, we find that
for $c<0$ this term has the same effect as adding a
$\overline{D3}$-brane in the proposal of KKLT. Including this
modification therefore provides a way of realizing the positive energy
contribution required for lifting an AdS vacuum to a meta-stable dS
vacuum. Indeed, we find that for $c<0$ the potential \eqref{1.4} has a
range of values of $A$ for which $z$ can be stabilized in a meta-stable
de Sitter vacuum.\footnote{For $c\geq 0$ this lifting does not work,
and we can stabilize $z$ in an AdS vacuum only.} This follows from the
observation that, as we show in the next section, the potential
precisely corresponds to the one analyzed in \cite{DSTV}, where details
and figures showing this dS vacuum can be found.

\section{Rigid Calabi-Yau compactifications} %%%%%%%%%%%%%%%%%%%%%%%%%%%

Ref.\ \cite{DSTV} studied perturbative and membrane instanton
corrections to the universal hypermultiplet arising from the
compactification of type IIA strings on a rigid CY$_3$. It was
shown that these quantum corrections can be described within the
framework of Przanowski-Tod metrics \cite{P,T}, i.e., the most general
four-dimensional QK metrics with (at least) one Killing
vector.\footnote{This Killing vector is present in the absence of
fivebrane instantons.} In local coordinates $r,u,v,t$ they
read\footnote{Note that, in contrast to \cite{DSTV} where $\kappa^2=
1/2$, we here work with the more standard conventions in which $\kappa^2
=1$. Compared to \cite{DSTV} this results in a rescaling of the scalar
field metric by a factor $1/2$, while the scalar potential is multiplied
by an overall factor of $1/4$. We refer to the appendix A of \cite{DSTV}
for details.}
 \begin{equation} \label{2.1}
  \d s^2 = \frac{1}{2 \, r^2} \Big[ f \d r^2 + f \e^h (\d u^2 + \d v^2)
  + f^{-1} (\d t + \Theta)^2 \Big]\ .
 \end{equation}
The metric is determined in terms of one scalar function $h(r,u,v)$,
which is subject to the three-dimensional Toda equation
 \begin{equation} \label{Toda}
  (\p_u^2 + \p_v^2) h + \p_r^2\, \e^h = 0\ .
 \end{equation}
The function $f(r,u,v)$ is related to $h$ through
 \begin{equation} \label{f-h}
  f = - \frac{3}{\Lambda} \big( 2 - r \p_r h \big)\ ,
 \end{equation}
while the 1-form $\Theta(r,u,v)=\Theta_r\d r+\Theta_u\d u+\Theta_v\d v$
is a solution to the equation
 \begin{equation} \label{dT}
  \d \Theta = (\p_u f\, \d v - \p_v f\, \d u) \wedge \d r
  + \p_r (f \e^h)\, \d u \wedge \d v\ .
 \end{equation}
The metric has a manifest isometry that acts as a shift of the
coordinate $t$. $\Lambda$ in \eqref{f-h} is the target space
cosmological constant; for the universal hypermultiplet we have
$R_{AB}=-3G_{AB}$, thus $\Lambda=-3$.

The coordinates $r,u,v,t$ can be identified with the fields in the
universal hypermultiplet coming from string theory via
 \begin{equation} \label{3.8}
  t=\sigma\ ,\qquad r = \e^\phi\ ,\qquad u = \chi\ ,\qquad v =
  \varphi\ .
 \end{equation}
Here, $\chi$ and $\varphi$ are the RR scalars originating from the
dimensional reduction of the ten-dimensional 3-form $\hat{C_3}$, $\phi$
is the dilaton, and $\sigma$ denotes the axion arising from the
dualization of the 4-dimensional part of the NS 2-form $\hat{B}_2$. With
this identification the universal hypermultiplet including the one-loop
correction of \cite{AMTV,ARV} is described by \cite{DSTV}
 \begin{equation} \label{pertm}
  \e^h = r + c \ ,\qquad f = \frac{r + 2c}{r + c}\ ,\qquad \Theta =
   u\, \d v\ .
 \end{equation}
For $c=0$ this solution gives the classical moduli space $\mathrm{SU}
(1,2)/\mathrm{U}(2)$ of the universal hypermultiplet. The one-loop
correction is determined through 
 \begin{equation}\label{defc}
  c = - \frac{4\, \zeta(2)\, \chi(X)}{(2\pi)^3} = - \frac{1}{6\pi}\,
  (h_{1,1} - h_{1,2})\ ,
 \end{equation}
where $h_{1,1}$ and $h_{1,2}$ are the Hodge numbers of the CY threefold
$X$ on which the type IIA string has been compactified. For rigid CY's,
where $h_{1,2}=0$, we have that $c<0$.\footnote{We expect that the
numerical value of $c$ obtained for a CY$_3$ orientifold differs from
this $N=2$ result, since projecting out states from the spectrum will
affect the resulting one-loop contribution. The sign of $c$, however,
should not change.}

Furthermore, it was shown in \cite{DSTV} that, upon including the leading
membrane instanton corrections, $\e^h$ takes the form
 \begin{align} \label{eh-exp}
  \e^h = r + c + \frac{1}{2}\, & r^{-m_1/2}\, \big( A_1\, \e^{\I v} +
	B_1\, \e^{-\I u} + \text{c.c.} \big)\, \times \notag \\[2pt]
  & \times \e^{-2\sqrt{r+c}} + \ldots\ ,
  \end{align}
where $m_1$ is an undetermined integer coefficient which has to satisfy
$m_1\ge -2$, and $A_1$, $B_1$ are complex constants. Ref.\ \cite{DSTV}
then explained that this expansion can be completed to a full solution
of the Toda equation \eqref{Toda} and that it reproduces the leading
order membrane instanton corrections to the four-hyperino couplings
predicted by string theory \cite{BBS}. Turning on a space-time filling 
$F_4$-form flux then induces a scalar potential in the
$N=2$ theory corresponding to gauging the shift symmetry of the axion.
Upon including the perturbative and leading order corrections in the
one-instanton sector, the resulting scalar potential reads \cite{DSTV}
 \begin{align} \label{2.1a}
  V = & \frac{e_0^2}{4\, r^2} - \frac{c\, e_0^2}{r^2 (r+2c)} - e_0^2\,
	r^{-(m_1+5)/2}\, \e^{-2\sqrt{r+c}}\, \times \notag \\*[2pt]
  & \times \re \big( A_1\, \e^{ \I v} + B_1\, \e^{-\I u} \big) +
	\ldots\ .
 \end{align}

\section{The orientifold projection} %%%%%%%%%%%%%%%%%%%%%%%%%%%%%%%%%%%

We now apply the orientifold projection to the universal hypermultiplet
discussed above, thereby truncating the theory to an $N=1$ supergravity
Lagrangian. According to \cite{GL}, this projection amounts to
eliminating the axion $\sigma$ and the RR scalar $u$. Using the tensor
multiplet description given in appendix B of \cite{DSTV}, the truncated
line element becomes
 \begin{equation} \label{2.2}
  \d s^2 = \frac{f}{2\, r^2} \big( \d r^2 +  \e^h\, \d v^2 \big)\ .
 \end{equation}
Substituting the solution \eqref{eh-exp} and taking into account the
leading instanton term only then yields
 \begin{align} \label{linefull}
  \d s^2 & = \frac{r+2c}{2 r^2\, (r+c)}\, \big[ \d r^2 + (r+c)\, \d v^2
	\big] \notag \\
    & \tab + \frac{1}{2}\, r^{-(m_1+5)/2}\, \re \big( A_1\, \e^{-2
	\sqrt{r+c} + \I v} \big) \big[ \d r^2 + r \, \d v^2 \big]
	\notag \\[2pt]
  & \tab + \ldots\ .
 \end{align} 
Here the first line arises from the perturbative corrections
\eqref{pertm}, while the second line displays the leading instanton
contribution. $N=1$ supergravity requires this metric to be K\"ahler.
At the perturbative level, this is easy to verify by defining the 
complex structures as in \eqref{zpert} and substituting it into the
metric
 \begin{equation}
  \d s^2 = 2\, \cK_{z\zb}\, \d z\, \d\zb\ ,
 \end{equation}
with K\"ahler potential given by \eqref{1.2}. If we now define the
complex structure
 \begin{equation}\label{non-pert-z}
  z = 2 \sqrt{r+c} + \I v - \frac{1}{2}\, \bar{A}_1\, r^{-(m_1+1)/2}\,
  \e^{-(2 \sqrt{r+c} + \I v)}\ ,
 \end{equation}
we find that the leading order one-instanton contribution in
\eqref{linefull} is \emph{still} captured by the K\"ahler potential
\eqref{1.2}. Hence, the K\"ahler potential receives perturbative
corrections stemming from the one-loop correction encoded by $c$ only;
the leading membrane instanton corrections can be absorbed in a
coordinate transformation.\footnote{We point out that the K\"ahler
potential \eqref{1.2} contains the one-loop corrections inherited from
the $N=2$ theory only. This does not exclude the possibility that the
$N=1$ theory gives rise to additional perturbative or higher order
non-perturbative corrections to $\cK$.}

The scalar potential \eqref{1.4} of the $N=1$ theory arises from setting
the $u$-dependent terms in \eqref{2.1a} to zero. At the perturbative
level, this is immediately clear after identifying $W_0=e_0$.

Comparing the leading order instanton correction is slightly more
involved, because one has to take into account the instanton correction
to the complex structure given by \eqref{non-pert-z}. As one can verify,
the leading order instanton correction to the complex structure
contributes to the potential \eqref{1.4} at subleading order only.
Hence, the potentials \eqref{1.4} and \eqref{2.1a} coincide upon
identifying
 \begin{equation} \label{2.3}
  A = -4\, e_0 A_1\ ,\qquad m_1=-2\ ,
 \end{equation}
i.e., if $m_1$ takes its lowest possible value allowed by the Toda
equation. This shows that the leading $N=2$ membrane instanton
correction provides a holomorphic contribution to the $N=1$
superpotential.

It would be interesting to see if the lifting mechanism discussed in
this letter also applies to more realistic orientifold models such as
the ones discussed in \cite{CFI} or compactifications on $G_2$ manifolds
\cite{A}.

\bigskip
\begin{acknowledgments} %%%%%%%%%%%%%%%%%%%%%%%%%%%%%%%%%%%%%%%%%%%%%%%%

We thank Thomas Grimm for providing the idea for this investigation, and
for carefully reading the manuscript. Furthermore, we thank Frederik
Denef and Marcos Mari\~no for stimulating discussion. UT was supported
by the DFG within the priority program SPP~1096 on string theory.

\end{acknowledgments}

\raggedright 

\end{document}